# Feedback and control systems for future linear colliders: White Paper for Snowmass 2021 Topical Group AF07-RF

D. Filippetto[1], C. Serrano[1], Q. Du[1], L. Doolittle[1], D. Wang[1], M. Bachtis[2], P. Musumeci[2], A. Scheinker[3], J. Power[4], M. Bellaveglia[5], A. Gallo[5], L. Piersanti[5]

[1] Lawrence Berkeley National Laboratory
[2] University of California, Los Angeles
[3] Los Alamos National Laboratory
[4] Argonne National Laboratory
[5] Italian National Institute of Nuclear Physics (LNF-INFN)

## 1. Executive Summary

Particle accelerators for high energy physics will generate TeV-scale particle beams in large, multi-Km size machines colliding high brightness beams at the interaction point [1-4]. The high luminosity in such machines is achieved by producing very small asymmetric beam size at the interaction point, with short durations to minimize beam-beam effects. Tuning energy, timing and position of the beam for optimal performance will require high-precision controls of amplitude and phase of high-frequency electromagnetic fields and real-time processing of complex algorithms. *The stability of the colliding beams has a large impact on the collider's effective luminosity. Therefore, the technology readiness level of diagnostic and control systems will be a major consideration in the collider design.* The technical requirements of such systems depend on the specifics of beam parameters, such as transverse and longitudinal dimensions, charge/pulse and beam pulse format, which are driven by the accelerating technology of choice. While feedback systems with single bunch position monitor resolution below 50 nm and latency <300 ns have been demonstrated in beam test facilities, many advanced collider concepts make use of higher repetition rates, brighter beams and higher accelerating frequencies, and will require better performance, up to 1-2 order of magnitude, demanding aggressive R&D to be able to deliver and maintain the targeted luminosity. For example, in order to preserve picometer-scale emittance values alignment tolerances at the nanometer level are needed throughout the accelerator.

Superconducting RF accelerators, such as the one envisioned for ILC [3], are highly sensitive to all perturbations, owing to the high quality factor of SRF cavities. In particular, mitigation of cavity detuning from microphonics and Lorentz force is still an area of active research. Advances in controls of SC cavities will greatly impact the total amount of RF power needed to operate the facility. Microphonics resonances exhibit sharp isolated peaks with frequencies up to 1 kHz, and require advanced complex controls (both electronic and mechanical) to counteract the cavity resonant frequency shift. In copper-based accelerators [1-2], the short RF pulse (<<500 ns) and temporal gap between consecutive electron beams (<< 10 ns) require new schemes for analog intrabunch beam-based feedback with latencies well below 100 ns, and advanced digital electronics for stabilizing RF amplitude and phase. Laser-based accelerator concepts [4] assume high-precision control of amplitude and phase of the laser pulse and sub-femtosecond synchronization between the laser and the incoming electron beam. In this case the compactness of the beam sensor element will also play an important role, as it will need to fit in between miniaturized consecutive acceleration modules.

At the same time, advances in computing hardware and architectures could be exploited to deal with the complexity of these large accelerator systems and to ease the tight requirements of new concepts. Non-linear optimization algorithms and comprehensive surrogate models of the entire machine can be implemented on modern fast computing architectures, such as GPU and FPGA boards, and interfaced in real-time with the machine, to contribute to the stability of electron and laser beams. Improvements in the precision of controls require adaptive controls that compensate for un-modeled time variation and disturbances. The need to handle time variation requires the combination of model-independent adaptive feedback, AI, and physics-informed models.

Efforts on developing specific accelerator LLRF hardware and software have historically been funded through single project developments, resulting in a disconnect among different US laboratories. Encouraging the use of commercially available components could be an effective way to reduce the effort in developing in-house electronics, and increase standardization. On the other hand the requirements in terms of speed and noise level necessary for large-scale facilities do not always align with industry needs, and the resulting system architecture may not be the optimal for the purpose of particle accelerator control. Standardization among research institutions should be pursued by creating a common development platform and through the use of open-source firmware and software, rather than trying to adapt all commercial solutions to accelerators. Furthermore, this field has many challenges in common with the technology of High Energy Physics Detectors. Such commonalities should be explored.

## 2. Introduction

The target luminosity of future advanced colliders ranges upward of 1E34 $cm^{-2}s^{-1}$ for linear designs. Such extreme performances will be achieved by means of extreme particle densities at the interaction point, and high average currents. Reaching the target luminosity requires extremely accurate control of beam parameters. For example, manipulation and transport of ultralow-emittance beams asks for nanometer alignment accuracy throughout the accelerator. Also, advanced technologies use very high accelerating frequencies to decrease the machine footprint to reach the desired center-of-mass energy. As the frequency increases, the dimensions, density and stability requirements of particle beams and accelerating units scale accordingly, and the accuracy and performance of diagnostic and control technology must follow. In two-beam accelerating schemes (either laser-particle or particle-particle beams) (sub-)femtosecond temporal synchronization between the driving and the main beam is key to reach collider-level stability, such as precise control of the drive beam parameters to obtain stable acceleration, (beam charge for wakefield acceleration, laser amplitude and phase in case of laser-driven acceleration). Similar synchronization requirements are needed in the case of high frequency cavities with very high gradients and short filling times. In this case, control and stabilization of the RF field amplitude and phase within the pulse is still very much an R&D topic, where feedback latency plays a major role.

Table 1 summarizes the machine parameters for different types of proposed linear collider technology. Although the development of advanced control will likely follow different paths depending on the machine type, several common elements can be identified. The tight requirements listed above call for developments of advanced non-disruptive diagnostics, fast feedback systems, high speed data acquisition and advanced edge-processing, assembled together in a modern architecture which optimizes loop latency and noise figures. Much of this is enabled by the tremendous progress of high speed digital electronics in the last few decades, mainly driven by telecommunication and security applications. On the other hand, some very low latency applications (<<200 ns) may be outside of the reach of digital processing, requiring all-analog circuitry.

The control system architecture of future linear colliders will have to provide the backbone infrastructure for realizing the full potential impact of modern computational tools in the control of particle beams (see Fig.1). The extended size of the machine will require precise timing distribution, a centralized computing unit (CCU), and local computing nodes for fast feedback. The CCU will run complex algorithms to continuously analyze the behavior of the whole system and create a dynamic virtual replica of the accelerator, Neural Network-based surrogate model of the machine. A static version of such a model will also run on the local nodes (RF stations including a precision receiver, signal processing and output control signal), and it will be updated periodically by the CCU. Time-coherent data will be sent to the individual nodes to generate the new setpoints and to the CCU, where a global Machine-Learning-based engine will perform continuous online re-training of the model. Such a scheme will allow efficient interception of correlated variations and slow drifts, performing a global optimization of the accelerator, managing variations of local feedback setpoints using an *holistic* approach. New computational methods can learn optimal feedback control laws for time-varying systems directly from data [5]. Fast local feedback nodes, such as FPGA-based RF field controllers, will use these setpoints to directly stabilize the single subsystems. They must be placed as close to the controlled components as possible to minimize signal delay because the feedback gain K of a simple proportional feedback controller of the form $dx(t)/dt = -Kx(t-D)$ is (to 1st order) limited to $\sim 1/D$, where D is the signal delay.

As machines grow in size and complexity, automated failure detection is also important for a centralized global control system communicating with multiple individual feedback nodes. As the number of high impact individual components grows, ML-based methods will become more important to predict or identify failures almost instantly and respond by shutting off devices to prevent damage or by adjusting nearby functioning devices to make up for the loss of a few individuals, which requires a global coordination between all of the components. For example, preliminary work towards ML-based automated fault detection includes SRF cavity fault classification at Jefferson Laboratory [6] and for BPM fault detection at CERN [7].

| | Units | CLIC 380 GeV | C$^3$ 250 GeV | ILC 500 GeV | AAC-LC* |
|---|---|---|---|---|---|
| **Accelerating field frequency** | GHz | 12 | 5.712 | 1.3 | 30-3000 |
| **Macropulse rep.rate** | Hz | 50 | 120 | 5 | 50000 |
| **Bunch spacing** | ns | 0.5 | 5.26 | 554 | 20000 |
| **Number of bunches in macropulse** | | 352 | 133 | 1312 | 1 |
| **Accelerating pulse duration** | µs | 0.244 | 0.7 | 1650 | $10^{-9}$-$10^{-2}$ |
| **Particles per bunch** | 1E9 | 5.2 | 6.24 | 20 | 5 |
| **Vertical beam size (beta) at IP** | nm (mm) | 2.9 (0.31) | 2.2 (0.12) | 5.9 (0.48) | 1 (0.5) |
| **Horizontal beam size (beta) at IP** | nm (mm) | 149 (18.4) | 148 (12) | 655 (11) | 25 (5) |
| **Normalized horizontal emittance** | µm | 0.9 | 0.9 | 0.5 | 0.87 |
| **Normalized vertical emittance** | nm | 20 | 20 | 35 | 20 |
| **Longitudinal bunch length at IP** | µm | 70 | 100 | 300 | 5 |
| **Energy stability** | % | 0.35 | 0.235 | 0.1 | 1.1 |
| **Energy spread** | % | 0.35 | 0.235 | 0.1 | 1.1 |
| **Accelerating field amplitude stability** | % | 0.1% | 0.1% | 0.07% | 1% ** |
| **Accelerating field phase stability** | Deg (fs) | 0.2 | 0.3 | 0.24 | (0.3-2) *** |

Table1: Beam parameters for different proposed linear collider technologies [1-4]; (*) Parameters are for a PWFA design, unless specified otherwise. Similar set of parameters exist for LWFA and SWFA; (**) Specifications are on beam energy asymmetry distribution. From [11]. (***) Reported values are for various schemes of LPA acceleration. From [11].

Laser-Plasma Accelerators (LPAs), advanced real-time computing techniques will be a key technology towards achieving higher beam quality. One promising technological path towards an LPA collider, is to coherently combine hundreds of parallelly amplified ultrafast fiber laser pulses temporally, spatially and spectrally, in order to meet the requirement of multi-kHz, high efficiency,

Joules of pulse energy and femtosecond pulse-width driving laser for each of ~100 LPA stages of a future collider [8], part of the US DOE's long term strategy. Preliminary results show the potential of deep learning models in helping to process the large amount of information for real time active coherence stabilization. Spatial combination of 81 channels has been achieved at LBNL using pre-trained deep learning models and model-free deep reinforcement online-learning [9-10].

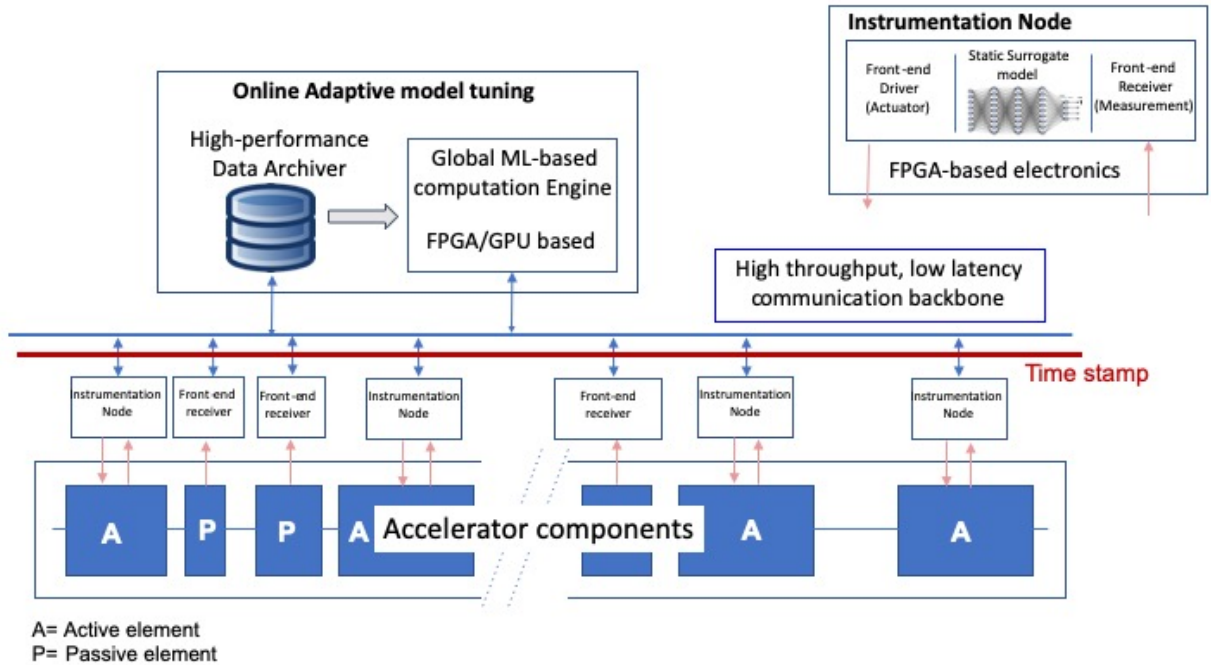

Figure 1: Possible scheme of a real-time control network of a particle accelerator.

## 3. Requirements

The control system requirements are generally separated into two different categories, 1) sensing and control applied directly to the beam, and 2) control of amplitude and phase of the accelerating electromagnetic fields. As mentioned above, the detailed requirements for a control system depend on the particular accelerator design. The technology used for acceleration vary from superconducting radio-frequency (SRF), high-peak power/high frequency RF cavities, or plasma from strongly ionized gas, each posing unique challenges in the control of amplitude and phase. In all cases though, the challenge includes a careful balancing of the main feedback parameters, i.e. latency, resolution in space and time (or amplitude and phase), and noise floor of the system. Below we review at a very high level the main present challenges for each of the major technologies.

## 3.1. Controls of Superconducting Accelerators

SRF cavities allow millisecond-long RF pulses, relaxing the requirements on feedback latency. The International Linear Collider (ILC) baseline proposal envisions thousands of bunches separated by ~500 ns in ms long bunch trains at repetition rates of 5 Hz. The ILC design is in an advanced state of technological readiness, with the main requirements for intra-train and inter-train control of beam parameters demonstrated in test facilities around the world [12].

Superconducting RF accelerators exhibit extremely high quality factors (~$10^9$-$10^{10}$), and consequently narrow resonance bandwidth which makes them highly sensitive to all perturbations. In particular, mitigation of cavity detuning from microphonics and Lorentz force is still an area of active research. Advances in controls of SC cavities will greatly impact the total amount of RF power needed to operate the facility [13]. Microphonics resonances exhibit sharp isolated peaks with frequencies up to 1 kHz, and require advanced complex controls (both electronic and mechanical) to counteract the cavity resonant frequency shift.

The narrow bandwidth of SRF cavities limits the speed of amplitude and phase adjustments relative to a fixed RF power source and their control could greatly benefit from predictive techniques [3,14]. Cold machines will likely benefit from the use of warm cavities with larger bandwidth both as sensors and actuators, to measure and compensate for high-frequency jitters in the beam [15].

## 3.2. Controls of copper-based RF accelerators

The extreme beam parameters proposed for advanced copper-based linear collider concepts [1-2] pose very stringent requirements on the accelerator diagnostic and control systems. Here the driving RF pulse length is limited to <1-to-few µs, at repetition rates of 50-120 Hz. A single RF pulse accelerates hundreds of bunches spaced by a few nanoseconds at most (see Table 1), pushing the required bandwidth of the feedback systems well beyond current technology.

Figure 2 shows an example of simulated feedback at the interaction point of the CLIC collider [1]. A total of 352 pulses are accelerated within a 12-GHz 244 ns-long RF pulse. The figure provides a clear view of the intra-bunch luminosity variation in the hypothesis of achieving a feedback latency of ~40 ns. Such value is one order of magnitude smaller than what is presently been experimentally achieved [16]. A potential route towards achieving similar performance has been traced via use of all-analog hardware [1], but its full implementation has not been demonstrated experimentally yet. Such an important problem would greatly benefit from a dedicated development, similar to what was done by the ILC collaboration at the test facilities, to increase its readiness level.

The short RF pulse duration requires active stabilization of field amplitude and phase of accelerating fields on the sub-microsecond level. Also, the transients generated by the beam passage in the accelerating unit require complex algorithms and fast LLRF systems to be compensated, minimizing detrimental effects on the beam and the amount of total RF power required. New digital boards, such as RadioFrequency Systems on Chip (RFSoC) boards, can achieve latencies in the sub-microsecond (Sect 4.2.3), but such systems are still not broadly used in the accelerator community, and validation of performance and characterization of the noise figure would be required to assess their potential.

*3.3. Controls for advanced accelerator concepts*

Collider designs based on laser or plasma acceleration techniques plan for equidistant pulses at tens of kHz repetition rates. Here more than the feedback response alone, it is the combination of sensitivity, compactness and latency that makes the overall diagnostic instrumentation challenging. Typical accelerating fields oscillate with frequencies in the range of tens of GHz (Structure-based Advanced Accelerators) to THz (PWFA and LPA) frequency. As the frequency increases the size of the actual accelerating stage scales accordingly, requiring sub-micrometer alignment between the injected electron beam and the electromagnetic center of the stage itself. For this class of accelerators, temporal synchronization between drive and main pulse is still a challenge. A time jitter of a fraction of the accelerating wave period induces a substantial change in output beam energy [11], requiring beyond state-of-the-art sub-femtosecond synchronization to stabilize it to better than 1%. A similar requirement holds between the drive and the main beam of Structure-based wakefield accelerators. In this case, the stability of the drive beam energy and time of arrival will directly impact the phase stability of the accelerating field experienced by the main beam.

Compactness of the sensor is a specification unique to these types of accelerators. Indeed in this case each electron beam diagnostic system could occupy a space along the accelerator equal or larger than the accelerating unit itself, lowering the effective integrated accelerating gradient.

Other specifications include precision control of driving laser energy and intensity distribution in LPAs, drive beam charge in PWFA, and plasma density fluctuations in both scenarios. In LPA, an asymmetry of 1% in the laser mode increases the emittance by one order of magnitude. Similarly, a plasma density variation of 1% will induce an equal amount of energy variation. More in general, for such small beams, obtaining a transverse jitter that is only a fraction of their size implies very tight requirements in the alignment of beam optics and the amount of acceptable mechanical vibration [11].

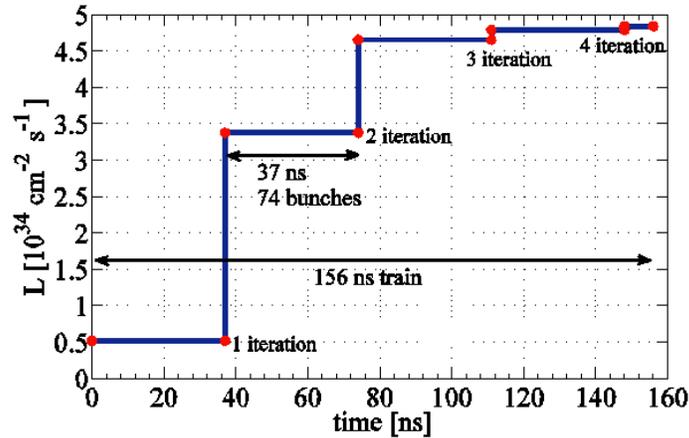

Figure 2: Luminosity enhancement with consecutive intra-train feedback corrections at the IP for a linear collider based on CLIC technology. From [17].

*3.4. Control system architecture*

The next generation of particle accelerators will have the opportunity to take advantage of novel tools developed in recent years thanks to the progress of high performance computing hardware and relative tools (see Fig.2 for example). In order to fully exploit such potential, novel control system architectures will need to be explored, where:

- Heterogeneous data (scalars, vectors, images, etc..) is archived at high speeds.
- A high precision phase reference is distributed throughout the accelerator. Data is distributed through deterministic links to all accelerator components, and time tagging is applied to all archived data to create a time coherent dataset.
- Data is distributed to one or more heterogeneous nodes (CPUs, GPUs, FPGAs, ASIC NN accelerators, etc…), for online and offline processing.
- Complex algorithms for adaptive automatic alignment procedures can be deployed and processed in real time.
- Edge-processed data should be exposed to the control system (e.g. EPICS) and become a 'first-class citizen' in the operation of the machine.
- The architecture provides access to a computing environment for real-time re-training of machine-learning models for prediction and virtual diagnostic, to be used for feedback and feedforward systems. Such computing capabilities should be leveraged to improve operational key parameters of the machines, such as global feedback performance and system reliability and calibration.

**4. Development path**

In terms of hardware needs, advancements on all aspects, including hardware, firmware and software are requested to meet the strict requirements of beam acceleration and control for the next generation of linear colliders. Measure and control of such beams will require judicious balance between conflicting requirements, such as bandwidth and sensitivity, and will include careful engineering of custom sensors and relative electronics, signal analysis/processing and archiving.

*4.1. Sensors*

Optimized design of electromagnetic sensors is essential to achieve the final requirement. It sets the limits on maximum signal-to-noise ratio and response time.
Feedback systems for stabilization of amplitude and phase require two probes of the same quantity to stabilize to function correctly. While the value of one probe is used in the control loop ("*in-loop*" probe) the other one is used as an independent measurement of the field in the cavity. When targeting very high stability (~1E-5) extreme care should be taken in making sure that the overall response, coupling and sensitivity of the two probes is identical. This includes cable shielding from EMI and minimization of crosstalk between different RF channels well below -100 dB.

Requirements on beam sensors include high sensitivity, linear response against beam parameters, and generation of fast signals enabling bunch-by-bunch discrimination.
One way of increasing the absolute sensitivity of diagnostic tools is to extend the working frequencies of EM pickups to higher frequencies, from C and X band, all the way to the optical domain. Also, the sensitivity could be increased by increasing the coupling between the sensor and the electron beam, i.e. by designing passive RF cavities with smaller irises. This choice would have an effect on the beam itself, and the wakefield budget would need to be taken in consideration.

The time resolution of the beam sensor is an important factor in the overall control feedback bandwidth, whose requirements depend on the particular acceleration method and pulse format used. The temporal response of the diagnostic becomes critical in the case of small spacing between consecutive bunches in a train, such as CLIC and $C^3$ and, when paired with tight spatial resolution requirements, it may call for dedicated R&D to reach the targeted specifications. As an example, in the case of the Beam Delivery System for CLIC, the simultaneous requirement for 10 ns temporal resolution and 3 nm spatial resolution are beyond the performance of present systems [16]. In order to increase the signal amplitude and decrease its duration, low Q-factor cavities are utilized for both measurement and control of beams [18]. On the other hand, the ultimate sensitivity of the sensor depends on its signal-to-noise ratio, which is ultimately limited by the Johnson-Nyquist noise. The integrated noise power increases with the measurement bandwidth (adding up to ~ -100dBm for 10 MHz, excluding the downstream electronic), correlating sensor speed and precision.

Non-destructive beam position and arrival monitors using THz/optical wavelengths should also be actively pursued, in particular in the context of optically driven accelerators, where direct mixing between beam-generated signal and the driving laser would provide the necessary resolution. Electro-optical sampling techniques have been demonstrated to be able to reach sub-30 fs resolution in time-of-arrival [19], so they provide a very promising research direction for non-destructive beam arrival time detection.

Lastly, it would be worth exploring the contribution that the mature field of nanotechnology and nanofabrication could provide to the control of ultradense electron pulses, developing sensors with enhanced coupling between beam and metal/dielectric surfaces.

## 4.2. Electronics

### 4.2.1. Analog intra-pulse feedback

To obtain ultimate low latency RF intra-pulse feedback, working to stabilize RF power stations, the most straightforward approach is to employ analog electronic components. In this case the latency is defined by the sum of the single component contributions. By selecting broadband RF circuitry (mixers, phase shifters, attenuators, wideband OPAMPs and ICs), a total loop response time below 100 ns can be pursued. Towards this goal, the latency of many sub-components of the feedback loop has been measured over the years, to provide a realistic estimate of the minimum loop delay achievable. Clearly the overall network length, including cable and printed circuits, plays an important role, such as the latency of the power sources in the loop, such as klystron group delay and/or other driving amplifiers.

As an example, the response time of a fast phase shifter working at 2856 MHz is reported in Fig.3. The phase shifter is driven by a TTL control step and shows a fall time of 6.3 ns (negative phase shift of about 180°). Fast signal processors [20] and high power drive amplifiers [21] have been tested at the ATF, demonstrating delays in line with the goal of sub-100 ns latencies, required to obtain a few intra-train corrections, as shown in Fig. 2.

A drawback of this kind of system with respect to their digital counterpart, is the larger sensitivity to external noise (EMI, RF crosstalk) and the difficulty in its customization once designed and produced. On the other hand, it is much cheaper and easier to realize and, as mentioned, it is expected to provide lower latencies.

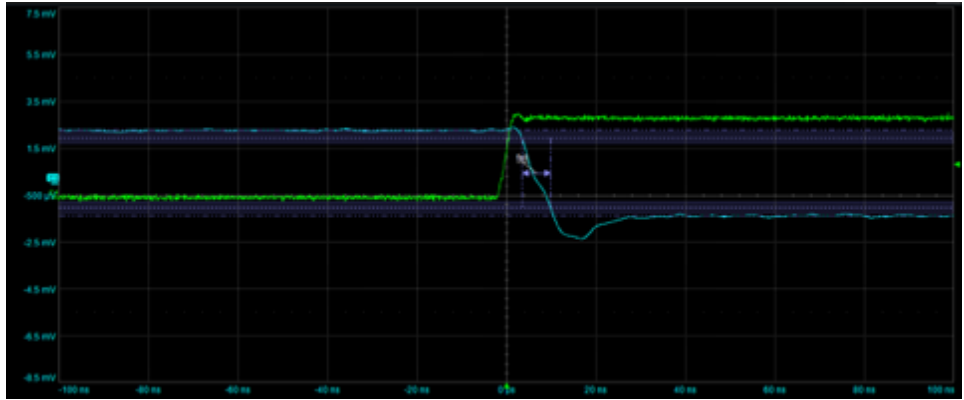

Figure 3: The response of a fast phase shifter (cyan waveform) reacting to a TTL step at the control port (green waveform). Phase shifter shows fall time of 6.3 ns (negative phase shift of about 180° at 2856 MHz). The short delay between the two curves is due to a slightly different cable length used to connect the signals to the scope.

*4.2.2. Digital Heterodyne processing*

Digital processing of RF data provides immunization from external sources of noise and allows complex operations of control and manipulation on chip. Signal detection using down-conversion techniques has demonstrated beam position detection resolutions below 30 nm, with 230 ns temporal response [16].
Radiofrequency detection and measurement is the first step in the signal chain of digital low-level-RF electronics (LLRF) and feedback systems [13][22], which can be seen as perturbance rejection mechanisms. The accuracy of the RF detection stage is not only the limiting factor for feedback systems, but also has a direct impact on accelerator construction and operational costs. For example, improving detection accuracy for the controls of accelerating cavities implies that the same level of stability can be achieved with lower levels of RF power, which in turns implies that smaller (and less costly) high-power amplifiers need to be procured at construction, and less RF power is consumed during operations, minimizing overall operational costs (and environmental footprint) of accelerator facilities [23].

The LCLS-II LLRF system [22] is a good example of state-of-the-art performance in RF controls of superconducting cavities, achieving RF stability of 0.01% in amplitude and 0.01 degrees in phase (RMS). With high-Q cavities, this is achieved by applying large feedback gains, extending the effective cavity bandwidth from 16 Hz to up to 40 kHz in the case of the LCLS-II SRF cavities, with a proportional gain of around 2400.

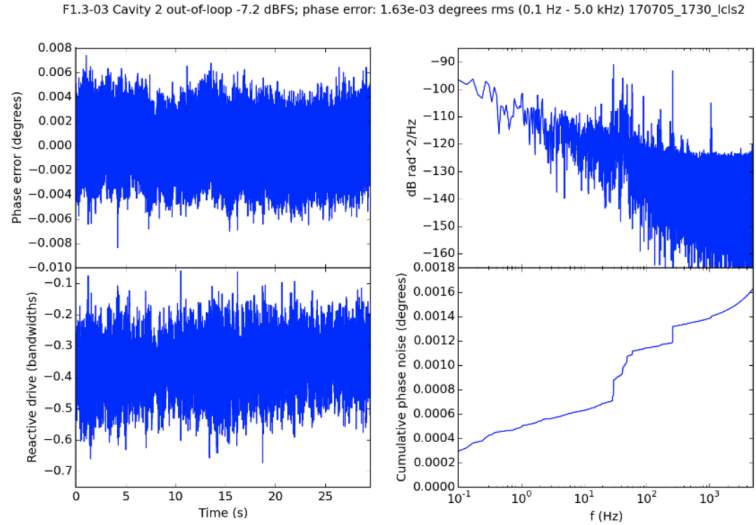

Figure 4: Measured phase noise of the LCLS-II LLRF system in a 1.3 GHz superconducting cavity.

Figure 4 shows the phase noise spectral density of the LCLS-II LLRF system, also the basis for other systems around the DOE complex such as PIP-II, where the phase noise is predominantly the 1/f noise component (-110 dBc/Hz at 1 Hz), vs the broadband noise component (-152 dBc/Hz, See Fig.5). The detection scheme used in this approach (and generally in LLRF systems) is heterodyne detection, where the RF signal of interest is mixed down to an intermediate frequency in the tens of MHz range. The objective of this approach is to use ADCs at around 100 MHz sampling rates, a regime where commercial ADCs have the best phase noise characteristics.

Another limit of the LLRF systems as currently engineered, with a 1/f component in the noise spectrum, is the need for periodic calibration to compensate for long-term drift (in LCLS-II specified as lower than 1 Hz) with invasive beam-based feedback techniques. This, along with overall RF measurement accuracy and its direct effect on the spatiotemporal resolution of pump-probe experiments, can be improved by addressing the fundamental limitation of the presence of the 1/f component in the measurement chain.

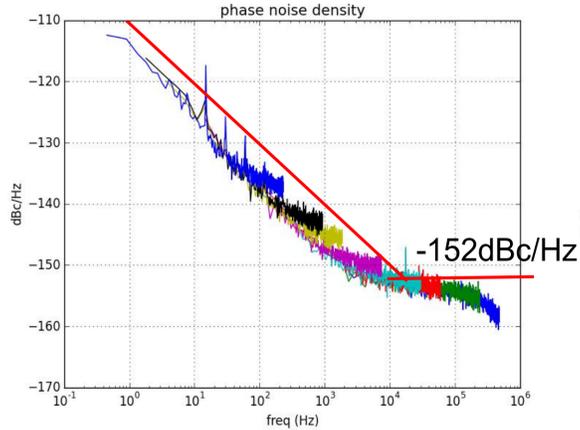

Figure 5: Power spectral density of a state-of-the-art RF measurement system.

The development of custom ADCs [24] could potentially greatly improve the noise figure of the system. For example, the addition of a chopping circuit to the ADC will allow to spectrally separate the desired digitized signal from the 1/f noise and subsequently filter out the 1/f noise, reducing the low-frequency ADC drifter. The chopper works as follows. First, the chopper modulates the input frequency to Nyquist (Fig. 6). Next, the ADC samples the chopped signal (Fig. 7). This will separate the desired signal from the 1/f noise spectrum. Then the resulting digital signal is chopped again (Fig. 8). This second level of chopping is done in the digital domain. Finally, the 1/f noise part of the noise spectrum is removed via low-pass filtering (Fig. 9).

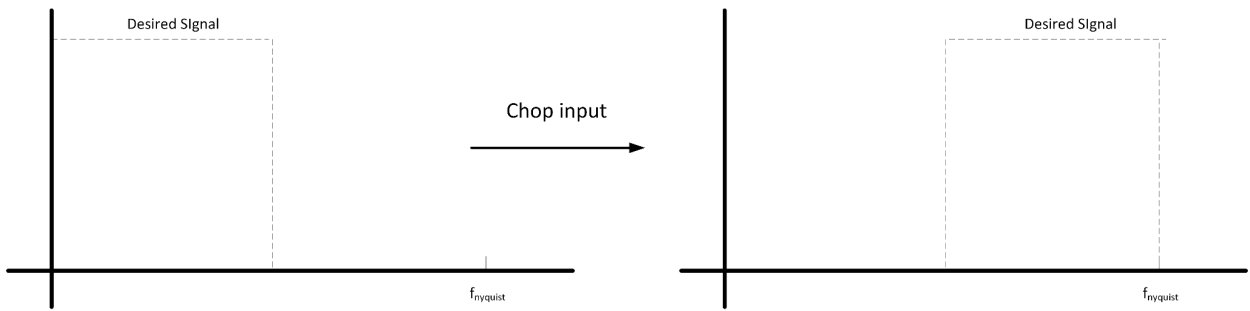

Figure 6. Chopping the input signal to modulate it to the Nyquist frequency.

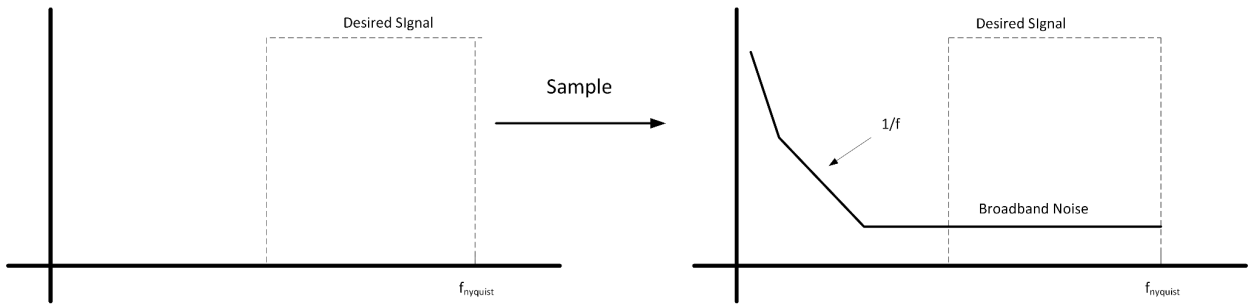

Figure 7. Sampling the signal superimposes the noise spectrum of the ADC on the signal spectrum.

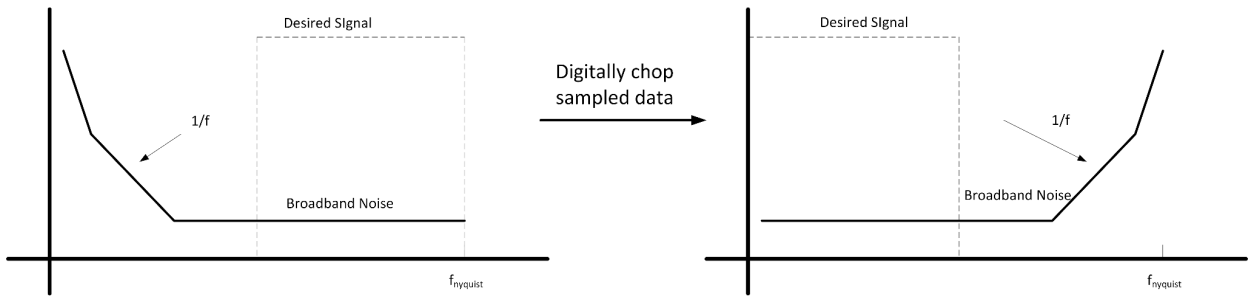

Figure 8. Digitally chopping the sampled data modulates the 1/f noise to Nyquist and returns the signal to baseband.

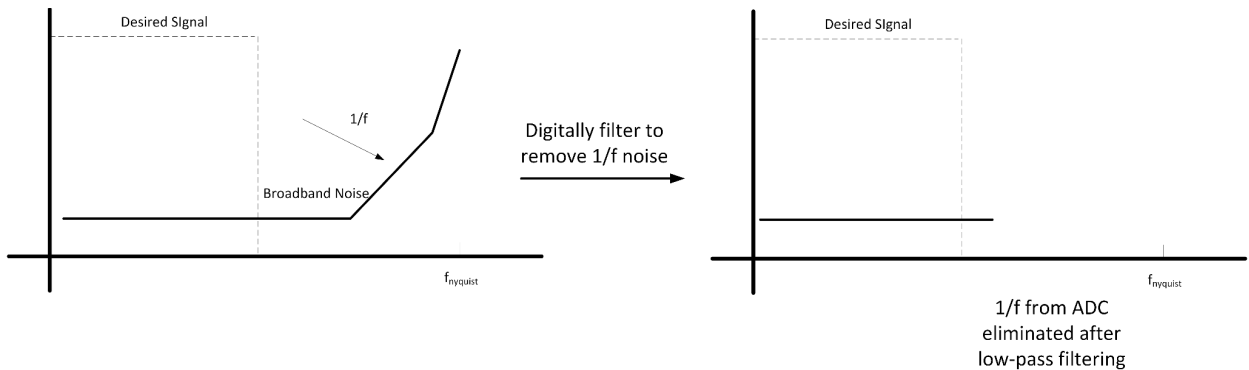

Figure 9. After digital chopping, the 1/f noise can be eliminated using low-pass filtering.

In addition, development of custom ADCs for physics applications will allow optimizing the broadband noise performance (partially by relaxing power consumption requirements) compared to commercial products, because we can focus on the noise figure while the commercial parts are designed to be general purpose and must balance multiple competing design considerations.

*4.2.3. Integrated RF converters in FPGAs and direct-sampling*

The use of integrated RF electronics with FPGAs within the silicon, capable of operating at high sampling rates, may be able to address one of the main limiting factors in current digital LLRF systems, latency, and can serve as a complementary technique to the heterodyne approach described above. Recent developments in 5G and radar applications feature low-latency Analog-To-Digital (ADC) and Digital-To-Analog (DAC) converters integrated together with FPGA logic and microprocessors in a single integrated circuit [25] . These devices reduce the latency by avoiding the digital data transmission from an ADC to an FPGA enabling intra-pulse or high-bandwidth feedback for amplitude and phase in RF structures. The ADC and DAC components in these devices support bandwidth up to 7 GHz and reach sampling rates up to 10 GSPS that enable direct sampling in S band and C band (second Nyquist zone) and traditional operation with an external I/Q modulator and mixer for higher frequencies. While the main gain of the approach is the lower latency, direct-sampling simplifies the system by requiring less external components and has also the benefit of avoiding the noise associated with the demodulation clock.

Preliminary studies on the capabilities of these devices have been performed in the Pegasus beamline at UCLA (Fig. 10) where the goal is intrapulse amplitude and phase stabilization on a high gradient S-band electron gun. The measured latency (~300 ns) could be further shortened by a factor of 2, by optimizing the algorithms as well as the geometry of the system to reduce cable lengths. This work will have to be extended to more complex systems and higher frequencies, with more extensive studies evaluating the performance in phase control, intra-pulse shaping. The printed-circuit-board (PCB) design for these mixed signal chips involves both RF and FPGA design in the same PCB and will be demonstrated towards compact future RF control systems for US Accelerator facilities.

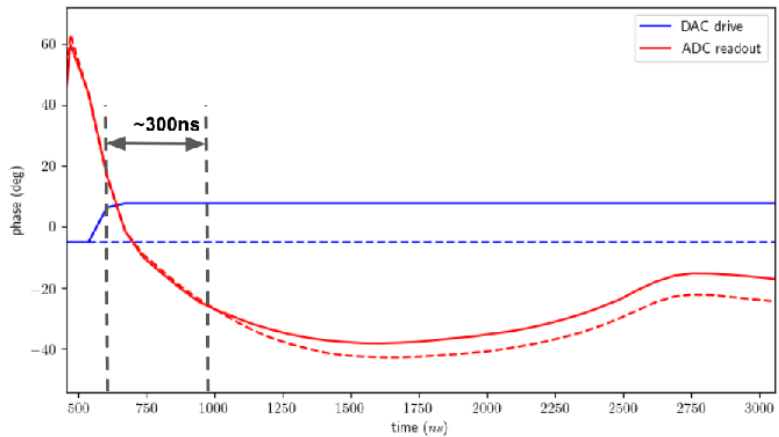

Figure 10: Demonstration of intra-pulse shaping in Pegasus ($f_{RF}$ = 2.856 GHz) showing an intrapulse variation of the output DAC and the resulting phase as read in input by the ADC (solid lines). Dotted lines correspond to a different pulse with no variation. The response time in phase change was estimated to be around 300 ns, including the propagation delay in the system.

*4.3. Enhancing controls with High Performance Computing tools*

Advanced computing techniques can help increase the stability and automatization of machine operations. THe use of complex algorithms substantially cuts down the optimization time, and can be used to cure drifts or avoid potentially dangerous situations. Digital twins of the real beamline can provide distributed virtual diagnostics along the accelerator, or enhance the resolution of real diagnostic tools by modeling noise patterns in the signal and performing real-time denoising.

The ability of solving complex non-linear multiparameter optimization in real time, to interface with the machine to apply corrections, and to learn from the response, has and will continue to enable breakthrough technology. An example is coherent beam-combined (CBC) lasers with precision optical control, a path to arbitrarily high laser energy and power. Multidimensional combination of ultrafast fiber lasers is a promising technique for building a Joule/kHz laser driver for LPA-based linear colliders [26]. Advanced controls are needed to facilitate laser operation and also optimize the entire system by sensing, diagnosing and providing end-to-end control. One popular solution called SPGD (stochastic parallel gradient descent) in CBC controls randomly dithers each input and measures the effect on the output without the deterministic knowledge of the system state, working its way up a multidimensional slope to find a point with zero gradient [27]. Although hundreds of CW channels with wide bandwidth have been coherently added this way, in the case of pulsed CBC the control bandwidth is limited by pulse repetition rate (1 kHz). Furthermore, the random dithering itself acts as a perturbation source, introducing excess noise. Using Data-driven ML-base algorithms turns out to be a powerful tool in error recognition in complex multi-input multi-output systems, and provides a robust technical path to the development of active feedback systems [29-30]. Figure 11 shows preliminary results in CBC feedback control, demonstrating real time active controls of CBC fiber laser as examples demonstrating utility to broad accelerator applications [30]. In the experiment, a neural network (NN) was trained to recognize system errors and feedback to correct system errors in a deterministic way. The comparison between the NN feedback and the traditional SPGD algorithms shows that the NN feedback has much better performance due to the accuracy of ML models.

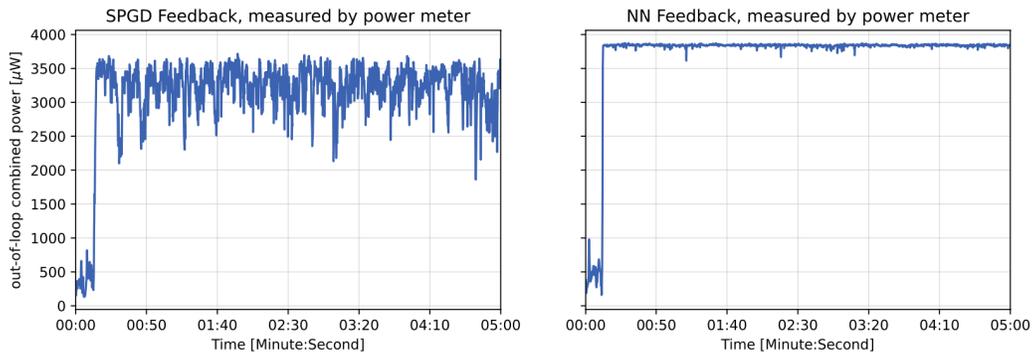

Figure 11: Experimental results from stabilization of spatial combining with 8 laser beams. Left, intensity stabilization based on SPGD algorithm; Right: replacing the SPGD with a NN-based feedback.

To perform online corrections using ML-based tools, the speed of processing is key to enable low latency feedback systems. Such need motivates the use of edge computers, i.e., fast-speed devices that process data on a local level. Possible edge computers include GPUs/Field-Programmable Gate Arrays (FPGAs) and Application Specific Integrated Circuits (ASICs). In particular, the FPGA, widely used in accelerators, is naturally amenable to ML models, can reduce the control latency of ML algorithms to microseconds and enable >1kHz repetition rate active control [31]. Such extreme performance will allow the possibility of implementing an online relearning procedure (i.e., repeat online the train of the ML periodically with new data). That capability will be profoundly beneficial for systems that need to be controlled over a long time or a system that, for some reason, changes its property over time. High-speed FPGA-based ML controllers are a new field and have a massive potential for advanced feedback systems in many applications.

Advanced feedback control algorithms can also be used to control the electron beam position, moments, and full charge distribution. Detailed control of the details of the beam phase space (both transverse and longitudinal) requires resolution beyond the existing state-of-the-art, Such accurate beam description can be provided by new virtual diagnostics tools, relative to which feedback can be applied. Various groups at accelerator facilities around the world have begun to develop ML-based tools to provide virtual diagnostics for accelerator operations. Researchers at the European XFEL at DESY have been developing incredibly high resolution non-invasive longitudinal phase space (LPS) diagnostics utilizing convolutional neural networks [32]. Researchers at SLAC National Accelerator Laboratory have also developed neural network-based virtual LPS diagnostics [33], and methods which utilize not only accelerator parameters as inputs, but also spectral measurements for increased resolution and prediction accuracy [34]. At CERN ML tools have also been developed as virtual diagnostics for not just the accelerated beam, but for the accelerator itself, for example for identifying magnet errors based on beam measurements [35]. At CERN surrogate models have also been developed for fast simulation studies of the CLIC final focus system, mapping sextupole offsets to luminosity and beam sizes without requiring computationally expensive tracking and beam-beam simulations [36]. AT PSI researchers have been utilizing advanced polynomial chaos techniques to develop surrogate models which utilize methods for modeling stochastic differential equations with uncertainty quantification which can be used to construct global sensitivity models with error propagation and error analysis [37].

One major challenge faced by existing and future accelerator facilities is time-varying components and beams. Because the initial phase space distributions of charged particle beams change as they are generated and enter the accelerator, and because accelerator components themselves drift with time, even if a perfect model could be made the correct initial conditions used as the beam input would be time-varying and uncertain. This problem of distribution shift is not special to accelerators, but is a general open problem in the ML community. There is a major need for adaptive machine learning (AML)-based algorithms which can be used to design diagnostics that can adapt quickly based on feedback and learned physics constraints without having to rely on new data acquisition for re-training. Such methods have recently shown the potential to provide predictions about the entire 6D phase space of intense charged particle beams with unknown and time-varying input beam distributions [38]. Such methods have also been used for predicting time-varying un-measured input beam distributions [39], in a first step towards AML-based diagnostics. An example of such application was demonstrated at the HiRES beamline at LBNL, in which an adaptively tuned convolutional neural network (CNN) coupled to an online adaptively tuned physics model was trained to map downstream beam images back to the associated time-varying initial transverse beam density $\rho(x,y)$ at the accelerator cathode as well as adaptively track time-varying accelerator components, as shown in Figure 12 .

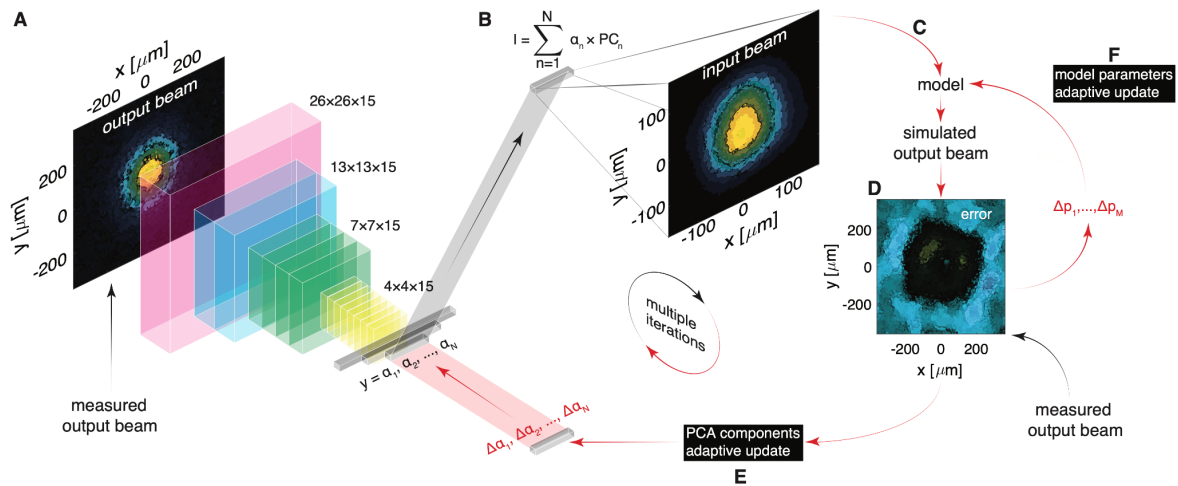

Fig. 12. The setup of an online model together with an adaptively tuned CNN for AML-based diagnostic of time-varying beams at HiRES is shown. A measured output beam (A) is mapped by the CNN to the principal components that generate the associated input beam distribution (B), which is then used as an input to an online model (C). The model's output is compared to the output beam measurement (D) and the error is used to adaptively tune both the CNN's output (E) and the online model (F) in order to track the time-varying input beam and time-varying accelerator parameters simultaneously. Figure adapted from [39].

## 5. Synergies with other applications

Production and delivery of ultrabright beams at high repetition rates is a common endeavor in many different user facilities across the DOE complex, beyond high energy physics applications. Fast and high precision sensing and feedback systems are therefore a key cross-cutting technology enabling the full potential of next generation large-scale instruments. Furthermore, the impact of the developments in high precision electronics will not be limited to particle accelerators, but will extend to the control of quantum computers, superconducting technology and quench detection, detectors and telecommunications.

The next generation of accelerators, including compact high repetition rate Ultrafast Electron diffraction (UED) setups and Free Electron Lasers, requires improvements in the precision of controls and adaptive controls that compensate for un-modeled time variation and disturbances. The need to handle time variation requires the combination of model-independent adaptive feedback, AI, and physics-informed models. A first of its kind adaptive ML-based fs-resolution longitudinal phase space control of the LCLS electron beam was recently demonstrated [40]. Such work needs to be extended to larger parameter range values and larger sets of coupled parameters for existing and future machines.